\documentclass[11pt,twoside,psfig]{article}

\def\farcs{\hbox{$.\!\!^{\prime\prime}$}}

\usepackage{asp2006}
\usepackage{epsf}
\usepackage{lscape}

\begin{document}
\title{Fabry-Perot Interferometry and Dynamics of Spiral Galaxies}   
\author{K. Fathi$^{1}$, J. E. Beckman$^{1}$, C. Carignan$^2$, O. Hernandez$^2$}   
\affil{
$^1$Instituto de Astrof\'\i sica de Canarias, La Laguna, Tenerife, Spain\\
$^2$LAE, Universit\'e de Montr\'eal, Montr\'eal, Qu\'ebec, Canada\\
}    

\begin{abstract} 
 We present two-dimensional Febry-Perot observations of emission-line distribution and kinematics in nearly spiral galaxies. We have developed and demonstrated the utility of a number of analysis tools which have general applicability, but which we have, so far, applied to only one galaxy (M~74, Fathi et al. 2007). In this galaxy, we have found kinematic signatures of radial motions caused by an m=2 perturbation. Such a perturbation may well be responsible for the inflow of material forming the nuclear ring and the inner rapidly rotating disc-like structure. The latter, in turn, could help build a pseudo-bulge. In the second paper in this series, we will apply the kinematic analysis tools to a sample of 9 late-type spiral galaxies observed with the FaNTOmM Fabry-Perot spectrometer at the Canada-France-Hawaii telescope.
 \end{abstract}



\section{Introduction}
In studying galactic dynamics, kinematic measurements over two spatial dimensions provide the necessary wealth of information. Currently used methods for achieving such observations, have either limited angular resolution (radio or submillimeter astronomy) or cover a relatively small field of view (integral field units). We use Fabry-Perot interferometry, which combined the high spatial sampling with large field of view, an ideal combination for studying extended objects such as galaxies. We aim to study a sample of 9 late-type spiral galaxies  
(NGC~3049, 
NGC~4294, 
NGC~4519, 
NGC~4654, 
NGC~5371, 
NGC~5921, 
NGC~5964, 
NGC~7479, 
NGC~7741), all observed with the FaNTOmM Fabry-Perot interferometer (Hernandez et al. 2003) at the 3.6 meter Canada-France-Hawaii telescope. The observations yiel the ideal combination of the spatial coverage and resolution by covering a field of $\sim 4^{\prime} \times 4^{\prime}$ with angular resolution of 0\farcs48/pixel. FaNTOmM is optimized for targeting the 6562.8 \AA\ H$\alpha$ emission-line, which we use to study the dynamics of the star-forming gas. 

\section{Methods}
The data are reduced homogenously, both following the standard data reduction packages and the improved IDL-based reduction package developed by Daigle et al. (2006). In Fig.~\ref{fig:N4519}, we illustrate the derived maps for one example  in our sample of 9 galaxies (NGC~4519). The velocity fields are used to analyze the rotational components by means of tilted-ring and harmonic decomposition techniques (e.g., Fathi et al. 2005). The rotation curve is in turn used to derive the angular frequencies and the effects of gravitational perturbations on the evolution of structures in the disk. In conjunction with HII region catalogues, we are then able to extract and quantify the kinematic effects of individual HII regions, and to separate these from the role of the gravitational perturbations on the observed velocity fields. Moreover, by averaging the velocity dispersion maps along elliptic annuli, we are able to derive the H$\alpha$-emitting gas velocity dispersion profiles to study the effect of feedback and disk heating mechanisms as a function of galactocentric radius. The data analysis methods have been applied on the nearby late-type spiral galaxy, M~74, and the utility of the codes are presented in detail in Fathi et al. (2007).
\setcounter{figure}{0}
\begin{figure}[!h]
\plotone{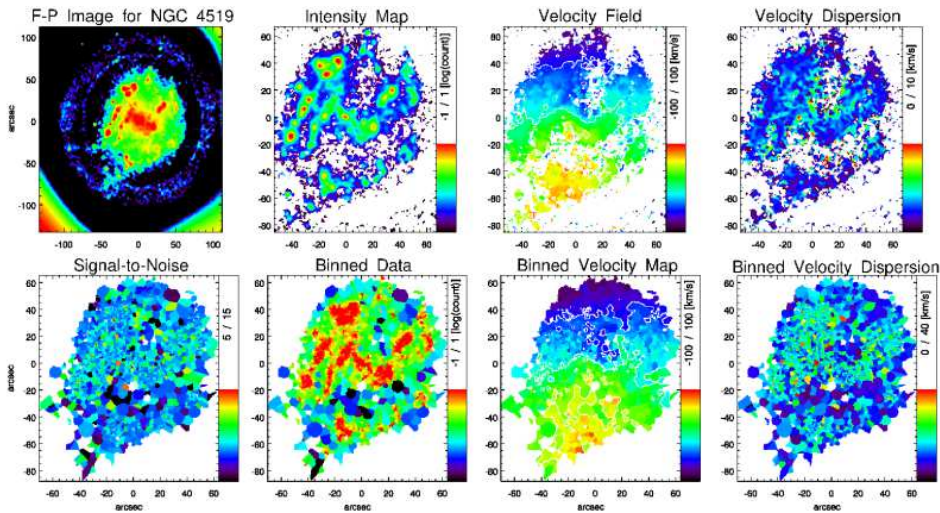}
\caption{Fabry-Perot H$\alpha$ observations of one of the 9 sample galaxies, NGC~4519. Illustrated are the results from the two different sets of reduction procedures. On the top row are the results from the reduction and derivation of the kinematics for all individual pixels, but only trusting the fits for pixels which have signal-to-noise larger than 3. The bottom row shows the result of the two-dimensional binning scheme, which increases the signal-to-noise, but reduces spatial resolution. The maps on the top row are ideal for studying individual HII regions, whereas the maps at the bottom row are optimized for analyzing the large-scale galaxy dynamics.} 
\label{fig:N4519} 
\end{figure}

\acknowledgements 
We thank the secretarial staff of the IAC, the LOC, and the SOC for an eclectic and  stimulating conference.

\end{document}